@begin definitions----------------------------------------
\documentstyle[12pt]{article}
\newfont{\bfit}{cmbxti10 scaled\magstep1}

\newcommand\beq{\begin{eqnarray}}
\newcommand\eeq{\end{eqnarray}}

\newcommand{\ee}{$(e^+e^-)\:$}
\newcommand{\eeee}{$(e^+e^+e^-e^-)\:$}

\newcommand{\r}{\cite}

\begin{document}

\begin{center}
  {\large\bf  The Doppler Paradigm and the APEX-EPOS-ORANGE 
Quandary\footnote{
This paper was presented at the XXVI Masurian Summer School of Physics in 
Piaski, Poland on Friday, September 1, 1996. It is available 
electronically from nucl-th@xxx.lanl.gov (or from 
nucl-th@babbage.sissa.it), or on the worldwide web\ at 
http://xxx.lanl.gov/ (or at http://babbage.sissa.it/) as
nucl-th/9601034, or in hard-copy from the author as U.of MD. PP\#96043.}}\\
\smallskip

\large  James J. Griffin \\
  Department of Physics, University of Maryland\\ \vspace{-.12in}
  College Park, Maryland, 20742, USA\\
   Internet: griffin@quark.umd.edu\\
\end{center}

\medskip

\centerline{\bf Abstract}

The experimental detection of the sharp lines of the \ee Puzzle is
viewed as a struggle against Doppler  broadening. Gedanken experiments
which are realistic in zeroth order of detail are analyzed to show that
the ORANGE and EPOS/I geometries select narrower slices of a Doppler
broadened line than spherically inclusive (APEX and EPOS/II --like)
apparati. Roughly speaking, the latter require event-by-event Doppler
reconstruction simply to regain an even footing with the former.  This
suggests  that  APEX' or EPOS/II's coincident pair distributions must
be statistically superior to those of EPOS/I or ORANGE in order to
support a comparable inference about sharp structure.  Under such
circumstances, independent alternative data is invaluable.  Therefore,
a corroboration of Sakai's 330.1 keV ($<$ 3 keV wide) electron line in
few MeV $e^+$ or $e^-$ bombardments of U and Th targets could prove
crucial.

\medskip
\noindent {\sl PACS Index Nos:12.20.Fv; 13.10.+q; 13.40.-f; 14.60.Cd; 
14.80.-j; 14.80.Pb; 23.20.Nx; 23.90.+w; 25.70.-z; 36.10.Dr }

\begin{center}
{\bf 1.~~ Introduction: Defeating Doppler Broadening of Sharp \ee Pairs} 
\end{center}

The most recent APEX\,\r{bett/95,chan/95} and EPOS/II\,\r{boke/95b}
U+Ta experiments report no evidence of the sharp $(e^+e^-)\:$ pairs
observed in earlier experiments of the EPOS/I
\,\r{cowa/85,sala/90,bar/95} and of the
ORANGE\,\r{bar/95,koen/89,koen/93} collaborations. We here view all of
the experimental searches for sharp pairs as incognizant attempts to
detect a set of sharp lines which have been Doppler broadened into a
smooth distribution. It is then possible to continue to believe {\it
all} the data, including the null results, and even to see in them
clarifications of old puzzles. This viewpoint also suggests the
selective culling of pairs as a fully parallel alternative to the use
of apparati designed to defeat Doppler broadening.

To explore this viewpoint, we utilize gedanken experiments which detect
the sharp rest frame energy of a uniform fully isotropic pair
distribution emitted from a moving source. These experiments are
analyzed for three experimental geometries.

The simplest ``Spherically Inclusive (SI)'' geometry resembles the APEX
and EPOS/II experiments; it measures an energy distribution which
exhibits the maximum possible Doppler broadening of the assumed
distribution.The second class of ``Bi-Hemispheric (BH) Selectors''
resembles the EPOS/I apparatus, and measures a distribution whose
Doppler width is reduced by its built-in selection of small values of
the total pair momentum, $P^\prime$.  The ``Angular Selectors (AS)'' of
our third geometry resemble the double-ORANGE apparatus and select a
narrower distribution by preferring pair momenta which are
perpendicular to the source velocity. It is remarkable that, from the
perspective ofthe Doppler Paradigm,the two apparati which have reported
sharp pairs emerge as well designed Doppler narrowers.

Under the guidance of these gedanken results, one learns that the
Spherically Inclusive (SI) apparati accept the broadest possible
distribution,and therefore begin the race one lap behind the
Bi-Hemispheric (BH) and Angular Selecting (AS) Doppler narrowers. By
applying event-by-event Doppler reconstruction to its pairs, an (SI)
machines which gathers complete pair information may largely overcome
this handicap. However, at the present level of precision, such
processing merely brings them up to rough parity with the selectors.

The question arises whether the bi-hemispheric data subset of the
spherically inclusive machines, enhanced by event-by-event
reconstruction, might provide the best prospect for improved data
quality.

As a by-product of this viewpoint there emerge also new insights, such
as the expectation that the ORANGE apparatus will detect underlying
sharp lines whatever the source velocity, but at energies modified by
an averaged Doppler shift of second order in the source velocity.

\begin{center}
{\bf 2. The Smoothing of Sharp Lines by Lorentz-Doppler Broadening}
\vspace{.12in}

{\bfit 2.1  The Uniform Isotropic $\Delta' = 0$   Data Distribution}
\end{center}

Consider a pristine \ee$ $ pair distribution which has, in the rest
frame of its source (wherein quantities are primed), a sharp sum energy
shared equally between the electron and positron, and a fully isotropic
distribution of pair sum and difference momenta. For specificity, we
take a source with velocity, {\boldmath $ v_S$} = 0.1c = {\boldmath
$\beta_S\,$c} $ $ in the beam direction, which emits pairs of sharp
total energy, ($\Sigma'\; = E_+' + E_-' = 3.48 \, {\rm mc}^2$,
corresponding to a pair  kinetic energy, $K_T' = 758 \, {\rm keV,}\;$)
and an average lab energy of $K_T$ = 750 keV. The isotropy implies also
that these pairs have opening angles distributed uniformly in the
range, $0\leq\;\cos\theta_{+-}^\prime \leq\;\pi$.  Then the quantity $(
\gamma \, \beta_S \,p') $ which occurs frequently has the value of
$\sim$ 70 keV.  Also, since $\beta_S$ is small and all of our leptons
are very relativistic, we take the lepton angles in the lab frame to be
equal to those in the rest frame of the source.

\begin{center}
{\bfit 2.2  Transformation to the Laboratory}
\end{center}

Note that in this distribution every pair has in the source rest frame
a sharp delta function energy and there is no ``background''. But the
Doppler-Lorentz tansformation to the lab frame spreads these sharp
energies and momenta into a Doppler broadened distribution.  Then a
pair whose total momentum is {\boldmath $P^\prime$} is observed in the
lab to have the energy,
 
\begin{equation} \Sigma  =  \gamma \Sigma' +
\gamma  \, \beta_S \,  P' \,\cos \theta_D^\prime =  \gamma \Sigma' +
D', \label{eq:A}\\ \end{equation} 

where $\theta_D'$ is the angle between the source velocity and the
total pair momentum, and   $D'$ is the ``Doppler Shift''. (We note that
for pairs with opening angles nearly equal to $180^\circ$, $P'$  is
small, and the Doppler broadening of the summed energy is not
important\r{grif/91a,grif/91b}, even if the decay occurs in a Coulomb
field with a finite positron/electron energy difference, such as EPOS
\r{sala/90}  observed in U + Ta.)

\begin{center}
{\bfit 2.3  Gedanken Experiment \#1: Spherically Inclusive Apparatus}
\end{center}

Consider an idealized ``Spherically Inclusive'' or ``(SI)'' apparatus,
which accepts all the pairs, and in particular all values of
$\cos\theta_D'$.  This apparatus will record a distribution of energies
for these pairs which has an average energy, $\Sigma^{SI}  = \gamma \,
\Sigma'$, (because $< \cos\theta^\prime > = 0\,$) and a width (defined
here as twice the rms deviation from the average), 

\begin{equation}
\Gamma^{SI} =  2 \gamma \beta_S \sqrt{<(P'^2)/3>} = 2 \gamma \, \beta_S
\, p' \, \sqrt{2/3}.    \label{eq:D} \end{equation} 

Here $<(P'^2)>$
denotes the average of $P'^2$ over the specified uniform opening angle
distribution, equal in this case to $2p'^2$, where $p'$ is the
magnitude of either lepton momentum, and $<\cos^2 \theta_D^\prime> =
1/3$.

(Note that for ``Spherically Inclusive'' apparati which accept equally
all pairs of the idealized isotropic data set, the width,
$\Gamma^{SI}$, is independent of the direction of the source velocity,
and therefore independent of the distribution of such directions. This
property does not extend to the symmetry breaking apparati discussed
below for which the assumed direction of {\boldmath $\beta_S$} along
the beam direction is therefore a nontrivial practical simplification.)

For the present (750 keV) example, the value of this width is
$\Gamma^{SI} \sim {\rm 120 \, keV}$. If similar considerations are
applied to an 805 keV line, then these two lines together yield in the
lab a smooth distribution in energy over the range from about 690 keV
to about 865 keV. Thus, the Sperically Inclusive apparatus will
evidence in its pair sum energy distribution no hint that the the  rest
frame energy distribution is composed of two sharp delta functions. The
extension to several lines, including perhaps some not yet identified,
is clear.

\begin{center}
{\bfit 2.4  SI Apparati Require Doppler Reconstruction}
\end{center}

If reasonable source velocities can Doppler broaden the sharp pair
decay lines beyond recognition in the energy distribution of the
spherically inclusive (SI) apparatus, how can such an apparatus verify
the underlying sharp lines experimentally? The answer lies in
supplementary measurements sufficient to define for each pair its own
Doppler shift,  $D^\prime$, and the utilization of these values to
carry out an event-by-event Doppler reconstruction of the rest frame
distribution.  In principal such complete information is now available
in the APEX and EPOS/II results, subject to some uncertainties. Then by
subtracting its own value of $D^\prime$ from the measured lab energy of
each pair, the rest frame energy (times $\gamma$) distribution could be
reconstructed.

For the APEX and EPOS/II experiments IPC pairs from $^{206}Pb$ have
been ``re-narrowed'' in this way to widths of the order of 40 keV
\,\r{chan/95,baum/95}.  This width is presumed to be set by
inaccuracies in the measured quantities comprising $D^\prime$ which
prevent the reconstruction from realizing the true value, zero.  We
take this 40 keV magnitude to indicate that such event-by-event
reconstruction can presently reduce the Doppler broadening of a pair by
a factor of about 40/120 = 1/3.

\begin{center}
{\bfit 2.5  Selection of the Unshifted Subset as an Alternative
to Doppler Reconstruction}
\end{center}

An alternative to the Doppler reconstruction is the selection of a
subset of pairs of small Doppler shift.  Equation~(\ref{eq:A}) states
that every pair for which the inequality,

\begin{equation}
 D' = \gamma  \, \beta_S \, P'  \, \cos\theta_D' \leq S, \label{eq:F}
\end{equation}

holds, will be measured in the lab to have an energy, $\Sigma$, which
is within S of ($\gamma$ times) the sharp rest frame energy, $\gamma
\Sigma ^{\prime}$. Here S is a pre-assigned limit chosen large enough
to provide a statistically potent subset. Then a culling process, which
retains  only the ``S-unshifted'' subset of pairs (with values of
$D'\leq S$),and which discards all other observed pairs, emerges as an
alternative to Doppler reconstruction.  The retained subset then
reflects the energy distribution in the source rest frame, and a study
of the subset may determine whether decay pairs of sharp energy occur
in the ion rest frame.  The answer  ``Yes'', would be indicated by
peaks in the number of ``S-unshifted'' pairs per unit sum energy
interval at ($\gamma $ times) the rest frame pair decay eigenenergies.

\begin{center}
{\bfit 2.6 Culling, or Mixed Culling/Reconstruction May Be
Advantageous in the  Presence of Large Backgrounds }
\end{center}
 
For the idealized no-background distribution of our gedanken
experiments, the Doppler reconstruction  would seem to promise a better
description of the rest frame energies than the S-subset culling
process. But in the actual experiments, the background is large, and
the signal is small.  Then the wholesale Doppler transformation of
background pairs required in the Doppler reconstruction might involve
subtle disadvantages, which the simpler process of culling could
perhaps mitigate.

Indeed, a mixed culling-reconstruction processes, in which both the
quality and the magnitude of the term $D^\prime$ are assessed in
deciding whether to discard the pair or to retain it for Doppler
reconstruction, might offer an optimal extraction of information in the
presence of large backgrounds.

We note also that the ORANGE and EPOS/I -like selector apparati
discussed below execute  automatic selections and rejections of pairs
by virtue of their very construction.  It seems difficult to argue that
this method of choosing a subset of pairs is intrinsically superior to
the culling process described above, especially when the culling
selections are guided by an intelligent physical question. The
essential question in the end is whether or not one discovers a true
physical property of the system.

\begin{center}
{\bf 3. Selecting Unbroadened Pairs by the Geometry of the Apparatus }
\end{center}

We next consider how an apparatus might be designed specifically to
select Doppler unshifted pairs from the laboratory distribution. The
basis for such a design is implicit in the ``unshifted'' pair selection
procedure outlined above: One seeks a design which selectively accepts
pairs with small values of the Doppler shift, $D^\prime$, of
eq.~(\ref{eq:F}), in order to achieve a measurement which is
informative of the rest frame pair energy distribution.

For a fixed value of the source velocity, small $D^\prime$ values could
be achieved by selecting small values of $\cos\theta_D'$ or small
values of $P^\prime$,  or both.  In fact, the culling procedure
described in sec. 2.5  above selects small values of the product of the
two, by minimizing the Doppler shift itself, pair by pair, and is
therefore manifestly optimal among such selections.

\begin{center}
{\bfit 3.1. Bi-hemispheric Selector Prefers Pairs with Smaller 
{\boldmath $P'$} Values}
\end{center}

A bi-hemispheric selector requires that of each pair, one lepton is
emitted into the Right (R) hemisphere and one into the Left (L), as in
the EPOS/I apparatus.  Apart from this requirement, the acceptance
range of electrons and positrons is broad in EPOS/I. In our idealized
bi-hemispheric selector, we assume it is perfect: every electron
(positron) emitted into to the Right (Left) hemisphere is accepted.

Then how does such a selector prefer small Doppler shifts,
$D^\prime$?  Obviously it accepts 1/2 of the pairs with opening angles
near $180^\circ$. (The other half yield leptons in the wrong
hemisphere.) But it accepts almost none of the pairs with opening angle
near $0^\circ$, since they almost always emit both leptons into the
same hemisphere. Then because the total sharp pair total momentum of
our idealized isotropic data set is given by

\begin{equation}
 P' =  p' \, \sqrt{2(1 + \cos\theta_{+-}')},    \label{eq:G}
\end{equation}

to select $\theta_{+-}'\sim 180^\circ$ is also to select values of $P'$
nearly equal to zero.  (For  more realistic data sets with $\Delta'
\neq 0$, the smallest $P'$ values are still selected by the $180^\circ$
preference, although they are not then equal to zero. See
Ref.\,\r{grif/91a}).

For intermediate $\theta_{+-}'$ angles, the fractional acceptance for a
pair with total momentum, P$^\prime$, with solenoidal (i.e.,w.r.t {\bf
B}-field) angles $\theta$  and $\phi$,  is given by the expression,

\begin{equation}
F(\theta_{+-};\theta) =
 {\rm ArcCos}[\cot(\theta_{+-}/2)\cot(\theta)]/(4\pi ^2). \label{eq:H}  
\end{equation}

for $\{(\pi/2\;-\;\theta_{+-}/2)\;\leq\;\theta\;\leq\;(\pi/2+
\theta_{+-}/2)\}$ and for $\{0\leq\;\phi\; \leq 2\pi\}$. It 
is zero for other  values of $\theta$.

With this weighting, we can calculate the width, $\Gamma^{BH}$ by
carrying out the average of ${D^\prime}^2$ for all pairs accepted from
our uniform isotropic distribution.  Again, we neglect the differences
between primed and unprimed angles in the first approximation.  For
each $\theta_{+-}$ value, the factor $\cos(\theta_D) =
\sin^2(\theta)\sin^2(\phi)$ must be averaged with the weight $F$ over
the angles $\theta$ and $\phi$ of the solenoidal coordinate system.
Then that result, multiplied by ${P^\prime}^2$ from (\ref{eq:G})
provides for each pair opening angle, $\theta_{+-}$, the squared
deviation of the energy from its unshifted value. Finally, the integral
of the squared deviation over the uniform distribution in
$\cos(\theta_{+-})$ of the uniform isotropic data distribution  yields
the following preliminary numerical estimate of the overall width of
the pair distribution accepted by the bi-hemispheric selector:

\begin{equation}
\Gamma^{BH} =  2 \gamma \, \beta_S \, p' \, \sqrt{(2)(0.08)} \sim 
0.5\Gamma^{SI},    \label{eq:I}
\end{equation}

Therefore, the BH geometry narrows the Doppler line by about a factor
of two, comparable to the factor of three given by Doppler
reconstruction of the IPC line in $^{206}Pb$.   (The average energy of
the distribution accepted by the BH selector is  given by the (SI)
value, $\gamma\,\Sigma^\prime$, because the BH average for
$\cos(\theta_D)$ again vanishes.)

\begin{center}
{\bfit 3.2. Angular Selector Prefers Small $\cos(\theta_D')$ Values}
\end{center}

We call an apparatus  an ``Angular Selector'' (AS) if it prefers small
$\cos\theta_D'$ values. It is therefore complementary to bi-hemispheric
selector discussed above, which prefers small magnitudes of $P'$. This
third gedanken geometry resembles the ORANGE apparatus, accepting
electrons on forward axial (about the beam) cones and positrons on
backward cones, which cones lie within some range of angles.  To
analyze this apparatus, we momentarily idealize it so that it accepts
only leptons emerging at a specific angle, $\theta_O$ with the beam
(instead of the range of angles, $\theta_O^{\scriptsize{MIN}} =
35^\circ \; {\rm to} \; \theta_O^{\scriptsize{MAX}} = 70^\circ$, which
the actual ORANGE detector accepts) and we imagine it to be in the rest
frame of the source instead of the lab frame. Then it would accept {\it
only} pairs from the uniform isotropic data set in which the total pair
momentum {\boldmath $P^{\prime}$} is perpendicular to the beam axis:
$\cos\theta_D'\;\equiv\;0.$  Thus, this slightly idealized ORANGE
apparatus is a {\it perfect} perpendicular selector, and therefore, a
perfect Doppler narrower.

The fact that ORANGE may actually accept an electron at $35^\circ$ and
a positron at $70^\circ$, and vice versa, and that it is fixed in the
lab, implies that some range of small but non-zero values for
$\cos\theta_D$  and for $\cos\theta_D'$ is actually accepted.  If the
accepted range is centered at $\cos\theta_D$ = 0, the $\cos\theta_D'$
will then  have a small average value $\sim \beta_S \, d'/P'$, where
the coefficient, $d'$ is defined by the average of $D^\prime$ over the
accepted distribution.
 Then our idealized isotropic data set would produce in this angular
 selector apparatus a distribution with the average energy,

\begin{equation}
\Sigma^{AS} =\gamma\Sigma'\;+\;\gamma d'(\beta_S)^2, \label{eq:K}
\end{equation}

(which exhibits a small Doppler shift, of second order in $\beta_S$,) 
and with a width due to the range of acceptance angles equal to
 
\begin{equation}
\Gamma^{AS} =\sqrt{2/3} \, \gamma \, \beta_S \, p
(\cos\theta_{\scriptsize{MAX}} - \cos\theta_{\scriptsize{MIN}}), 
\label{eq:L}
\end{equation}

This (AS) width is smaller than the width, (Eq.(2), accepted by the
idealized spherically inclusive (SI) apparatus by the factor,

\begin{equation}
\Gamma^{AS}/\Gamma^{SI} = (\cos\theta_O^{\scriptsize{MAX}}
-\cos\theta_O^{\scriptsize{MIN}})/2 \sim 0.25. 
\end{equation}

The Angular Selector therefore narrows the line by a factor of about
1/4 from that measured by the Spherically Inclusive apparatus. This
factor is to be compared with the factor, 1/2, for the Hemispheric
Selector.  Recall also that when the data of the Spherically Inclusive
selector includes the information required for Doppler reconstruction,
that process narrows the width by a factor of 1/3.

 We thus arrive at the conclusion that in this zeroth order of
 complexity, the Spherically  Inclusive (APEX and EPOS/II -like)
 apparatus is more blinded by Doppler broadening than the (EPOS/I and
 ORANGE -like) selectors, and that it  acquires a comparable
 sensitivity only after  event-by-event the Doppler reconstruction of
 its data.

\begin{center} 
{\bfit 3.3 New Insights Into the ORANGE--EPOS/I Results} 
\end{center}

This view of the the ORANGE apparatus as a suppressor of Doppler
broadening also enlightens our view of its measured results, and of
their relationship to the EPOS/I data by showing us that

(a) the ORANGE peak average energies ought not (NOT!) to be the same as
those of EPOS. Rather, they should differ by a small (second order in
$\beta_S$) averaged Doppler shift, whose magnitude reflects the
distribution accepted by the apparatus.  Therefore, the differences
between sharp energies of reported by the EPOS/I and the ORANGE
collaborations should be viewed not as measurement errors, but as
beckoning experimental hints;

(b) the ORANGE apparatus, presented with a uniform isotropic pair
distribution from a moving source, will record peaks corresponding to
energy delta functions of the rest frame for any magnitude of the
source velocity, and not just for sources at rest in the heavy ion
center of mass frame.

(c) Furthermore, in principle, the ORANGE experiment, at the cost of
diminishing the counting rate, could, by simply decreasing the angular
range of its acceptance cones, reduce the measured widths of its lines
to the lowest value allowed by the experimental conditions.
 Thus when, viewed as a gedanken anti-Doppler experiment, it has the
 remarkable property of full adjustability of the width-acceptance: any
desired narrowing can be achieved.

It is very important that the inference (b) above seems to conflict
with the repeatedly published assertions of the ORANGE group that their
source is nearly at rest in the center of mass frame and that it could
not possibly be moving with the velocity of the emerging
projectile-like ion\,\r{koen/93,koen/92}.  The present author has
however not found in the literature any specific supporting argument or
evidence for these assertions. It would be appropriate for the ORANGE
group to publish the detailed reasons for these claims, so that a
review of their validity could be carried out, and a definitive
conclusion reached whether any empirical evidence truly restricts the
velocity of the source of the sharp U+Ta \ee lines.

\begin{center} 
{\bf 4. Sakai's Alternative Sharp Lepton Evidence Needs Corroboration} 
\end{center}

Very sharp ($\Gamma\leq 2.1\; keV$) 330.1 keV electrons have been
reported repeatedly\,\r{saka/88,saka/91,saka/92a,saka/93} by Sakai, et
al. to emerge from irradiations of U and Th with energetic
$\beta^+$-decay positrons. Furthermore, a scenario in which these are
the electron partners of \ee pairs emitted from the same source as
provides the pairs of the  heavy ion ``$(e^+e^-)$--Puzzle'' has been
suggested\,\r{grif/94,grif/95a}. To test this connection it is
essential to understand these sharp leptons better.  One major
experimental gap is the absence of studies with sharp electron and/or
positron beams to corroborate Sakai et al.'s $\beta$-decay positron
studies.

If Sakai's data are true, they augment the $(e^+e^-)$--Puzzle of the
heavy ion  processes by providing invaluable evidence from an entirely
independent direction. Before the very recent null results of EPOS/II
became  known, it had been our intention to devote this paper  to the
important question of those sharp leptons, and to urge experimental
efforts to check Sakai's work.

Instead, we simply note here that our analysis of the Sakai data
suggests\,\r{grif/95d} that experiments with beams of 2 to 4 MeV
electrons or positrons should expect to measure the sharp 330.1 keV
electron and/or positron of the pair with a cross section of $10^2$
mb.  In addition, the scenario recommends (but in this case without a
good estimate of the cross section)  a search for these same sharp
electrons and positrons with positron beams in the 660 to 795 keV
resonance absorption range impinging upon U and Th atoms.

\begin{center}
{\bf 5. Relevance of  the Composite Particle $Q_0$ Scenario}
\end{center}

For several years  the author has considered a composite particle
creation/decay scenario to be the simplest framework capable of
encompassing all the data of the heavy ion ``\ee Puzzle''
\,\r{grif/91a,grif/88c,grif/89ae}. About the internal structure of the
composite particle, the data so far says nothing,
 but Occam's razor prefers a bound Quadronium \eeee
atom-without-a-nucleus, as the soundest present choice:  not
inadequately simple, but invoking no unecessary new hypothesis.  We
therefore refer to the scenario as the ``Composite Particle ($Q_0$?)
Scenario''.

The Composite  Particle ($Q_0$?)Scenario allows for spontaneous
Landau-Zener creation of $Q_0$ from the vacuum in strong enough heavy
ion Coulomb fields\,\r{grif/88b}.  It also provides a good
semiquantitative description\,\r{grif/91b} of EPOS' observed sum and
difference widths as arising from Doppler and Coulomb broadening.
Thus, the central assumption of the $Q_0$/EPOS phenomenology that
Doppler spreading  defines the observed sum widths, augmented by the
recent ORANGE evidence\,\r{koen/93} that these pairs frequently have
small opening angles,  leads directly to the main idea of this paper:
that Doppler spreading is the curtain which obscures the sharp decay
lines of the $Q_0$ eigenstates.  The scenario also  allows
for\,\r{grif/91c} exotic decays of $Q_0$ bound in a supercomposite
molecule  with a nuclear ion, emitting $(e^-e^+e^+), \; (e^+e^+), \;
(e^+)$, and one-$\gamma$, which have not yet been observed. Of these
the single $e^+$ Sharp Annihilative Positron Emission
(SAPosE)\,\r{grif/89d} presents an inverse creation process, Recoilless
Resonant Positron Absorption (RRePosA)\,\r{grif/94}, which may have
been observed in Sakai's studies mentioned above.

Another of these $Q_0$ bound supercomposite decays---the  annihilation
into one-$\gamma$---may already be in evidence as the 1780 keV gamma
ray commonly attributed to Uranium nucleus: if the IPC or EPC continue
not to suffice, then the observed gamma may just be an alternative
decay product of the bound $Q_0$.

The C($Q_0?$) Scenario survives the claim\,\r{tser/91} that all such
composite particles are excluded by high precision Bhabha results.  The
argument is based upon assumptions, not
measurements\,\r{grif/93a,wu/92,grif/93b}.  So far one can say that
$Q_0$, if it exists, must be strange, and that it might offend our
intuitions, but never that it contradicts known physics.

\begin{center}
{\bf 6. Summary and Conclusions}
\end{center}

When the experimental efforts of the ``\ee Puzzle'' are viewed as an
incognizant struggle to detect the sharp lines underlying a smooth
Doppler broadened distribution, then the ORANGE and EPOS/I apparati are
recognized as excellent ``Doppler Narrowers'', while the APEX and
EPOS/II effect no narrowing at all.  Gedanken experiments on an
idealized uniform isotropic pair data set exhibit these differences.

In this zeroth order of complexity no very narrow lines are selected
from the uniform isotropic distribution by any of the apparati, leaving
the explanation for the sharp ORANGE-EPOS/I lines to be sought in more
specific properties of the actual distributions and/or of the actual
apparati.  Also the inferiority of the APEX-EPOS/II geometry is brought
to parity with the selectors, but not to clear superiority, by
event-by-event Doppler reconstruction based on the detailed
supplementary data  for every pair. Thus  the benchmarks of this
Doppler Paradigm suggest that data from the APEX-EPOS/II geometries
must be substantially superior statistically to ORANGE-EPOS/I data
merely to match their Doppler narrowing capacities.

This viewpoint suggests also that differences between the EPOS' and
ORANGE's measured sharp energies must be expected, and that they should
be attributed to a second-order Doppler shift which reflects the
accepted distribution.

The Doppler Paradigm's proposed view of the ($e^+e^-$)--Puzzle problem
hinges upon its assumption that the sharp U+Ta lines originate in a
source moving with an
 emergent ion velocity. It therefore directly confronts the repeated
assertions\,\r{koen/93,koen/92} of the ORANGE group that this
assumption is excluded by their data.

In an ambiguous situation such we now confront, independent alternative
data can serve a crucial role.  We note that Sakai, et al.  have
repeatedly reported measuring (very!) sharp electrons  from
bombardments of heavy atoms by beta decay positrons.  We urge that
those studies be checked by analogous experiments with beams of few MeV
electrons or positrons.

The author is grateful to  H. Bokemeyer and  T. E.  Cowan for helpful
conversations and suggestions.  The support of the U.  S. Department of
Energy under grant no. DE-FG02-93ER-40762 is also acknowledged.


\begin{thebibliography}{99}

\bibitem{bett/95}
I. Ahmad,  et~al.,
\newblock {\it Nucl.Phys.A}{\bf 583}, 247(1995).

\bibitem{chan/95}
K.~C. Chan, 
\newblock {\em A Search for Correlated Positron-Electron Emission.},
\newblock PhD thesis, Yale University, 1995.

\bibitem{boke/95b}
H. Bokemeyer, 
\newblock (Private Communication.) 

\bibitem{cowa/85}
T. Cowan, H. Backe,  et~al.,
\newblock {\it Phys.Rev.Lett.}{\bf 54}, 1761(1985).

\bibitem{sala/90}
P. Salabura,  et~al.,
\newblock {\it Phys.Lett.}{\bf B245}, 153(1990).
\newblock See also references therein. 

\bibitem{bar/95}
 R. Baer, et~al.,
\newblock {\it Nucl.Phys.}{\bf A583},  237(1995).


\bibitem{koen/89}
W. Koenig, et~al.,
\newblock {\it Phys.Lett.}{\bf B218}, 12(1989).

\bibitem{koen/93}
 I. Koenig, et~al.,
\newblock {\it Z.Phys.}{\bf A346},  153(1993).


\bibitem{baum/95}
J.~Baumann, et al.,
\newblock in {\em 1994 GSI Laboratory Report No.GSI 95-1}, 
GSI Darmstadt, Germany.

\bibitem{grif/91a}
 J.~J. Griffin,
\newblock {\it Intl.J.Mod.Phys.}{\bf A6},  1985(1991).

\bibitem{koen/92}
W. Koenig and H.Tsertos,
\newblock {\it Phys.Rev.Lett.}{\bf 68},  1959(1992).

\bibitem{saka/88}
 M. Sakai, et~al.,
\newblock {\it Phys.Rev.}{\bf C38},  1971(1988).

\bibitem{saka/91}
 M.  Sakai, et~al.,
\newblock {\it Phys.Rev.}{\bf C44}, R944(1991).

\bibitem{saka/92a}
M. Sakai, et~al.,
\newblock in {\em Nuclear Physics of Our Times} 
(Sanibel Island, Florida, Nov.
  1992), edited by A. V. Ramayya, World Scientific Co.,
  Singapore, pp. 313--321.
.

\bibitem{saka/93}
M. Sakai, et~al.,
\newblock {\it Phys.Rev.} {\bf C47},  1595(1993).

\bibitem{grif/94}
 J.~J. Griffin,
\newblock in {\em Topics in Atomic and Nuclear Collisions}, 
Predeal Romania, Sept. 1992, ed. A. Calboreanu, V. Zoran, 
Plenum Press, 1994, page p419.

\bibitem{grif/95a}
 J.~J. Griffin,
\newblock in {\em Nucleus-Nucleus Collisions V}, Taormina, 
Italy, May 1994, ed M. DiToro, E. Migneco, P. Piatelli, 
North Holland, 1995, page p257.

\bibitem{grif/95d}
J.~J. Griffin,
\newblock Workshop on Positron Collisions, July 1995,
\newblock University of British Columbia, Vancouver, B.C.,
\newblock to be published in Can.J.Phys.{\bf 74}, XXX(1995).

\bibitem{grif/88c}
J.~J. Griffin,
\newblock in {\em Proc. 5th Int. Conf. on Nuclear Reaction 
Mechanisms}, Varenna, Italy, June 1988, edited by Gadioli, E.,  
(Physics, U. of Milano, Italia) page 669.

\bibitem{grif/89ae}
 J.~J. Griffin,
\newblock {\it J.Phys.Soc.Jpn.}{\bf 58},  S427(1989),
\newblock and earlier refs. cited there.

\bibitem{grif/88b}
J.~J. Griffin,
\newblock ``Quadronium: Spontaneous Creation in a Strong 
Coulomb Field'', U.Md.PP 89-053, (DOE/ER/40322-052) Oct. 1988.

\bibitem{grif/91b}
 J.~J. Griffin,
\newblock {\it Phys.Rev.Lett.}{\bf 66},  1426(1991).

\bibitem{grif/91c}
 J.~J. Griffin, 
\newblock in {\em Proc. 6th Int. Conf. on Nuclear Reaction 
Mechanisms} Varenna, Italy, June 1991, ed. E.Gadioli, Physics, 
U. of Milano, Italia, page 758.

\bibitem{grif/89d}
J.~J. Griffin and  T.~E. Cowan,
\newblock ``Prediction of Sharp Annihilative Positron Emissions 
from U+Ta Reaction''; U.Md.PP 90-060 (DOE/ER/40322-088) Nov. 1989.

\bibitem{tser/91}
 H. Tsertos, et~al., 
\newblock {\it Phys.Lett.}{\bf B266},  259(1991).


\bibitem{grif/93a}
J.~J. Griffin,
\newblock {\it Phys.Rev.C}{\bf 47},  351(1993).

\bibitem{wu/92}
X. Y. Wu et~al.,
\newblock {\it Phys.Rev.Lett.}{\bf 69},  1729(1992).

\bibitem{grif/93b}
 J.~J. Griffin,
\newblock {\it Phys.Rev.Lett.}{\bf 70},  4158(1993).

\end{thebibliography}
\end{document}